# ROTATING-SLIT SCINTIGRAPHY USING SCINTILLATING GLASS FIBERS: FIRST RESULTS

P. Ottonello[1], P. Pavan[2], G. Rottigni[1], G. Zanella[2] and R. Zannoni[2]

[1]*Dipartimento di Fisica dell'Università di Genova and I.N.F.N.-Sezione di Genova, via Dodecaneso, 33, 16146 Genova, Italy*
[2]*Dipartimento di Fisica dell'Università di Padova and I.N.F.N.-Sezione di Padova, via Marzolo, 8, 35131 Padova, Italy*

## Abstract

In this paper we propose to perform the scintigraphy of small organ using a rotating-slit collimator and a bundle of scintillating glass fibers, put in parallel with the slit and rotating with it. An intensified CCD, coupled to the end of the fibers, acquires an integrated image of the events per each rotation angle. The final image is computed by a back-projection procedure.
The advantages of this method, with respect to conventional scintigraphy, are the improvement of the detection efficiency of one-two order of magnitude without counting rate limitations, the improvement of the spatial resolution, the elimination of the parallax error and the rejection of the spurious events, without energy analysis. Simulations and first experimental results are showed.

## 1. Introduction

Scintigraphy is a common diagnostic technique both in nuclear medicine, and in the study of the radiopharmaceuticals on small animals in laboratory. In practice, *it measures the distribution (spatial and temporal) of a radioisotope, gained by the portion of the organism of which is wanted to underline the functional activity*. The radioisotope is administered through a radiopharmaceutical characterized by a preferential absorption in the zone of interest. The radioisotope mostly employed is Technetium-99m (99mTc) which decays in 99Tc, with emission of a single photon of 143 keV, with half-life of about 6 hours. In comparison to the PET, scintigraphy allows to follow kinetic processes for some hours, differently from the tens of minutes allowed from the radioisotopes used in the PET, so resulting complementary the two methods. Besides, scintigraphy, planar or tomographic, is used for the facility of production of 99mTc and the possibility to obtain radiopharmaceuticals of wide distribution.

The classical detector used in scintigraphy is the gamma-camera, evolution of the Anger camera [1]. It is generally constituted by a parallel-hole collimator coupled to a bulk scintillator, generally NaI(Tl). Unfortunately this collimator limits the angle of acceptance of radiation to be received by the scintillation crystal. The read-out is performed by a set of photomultipliers (PMTs), followed by an high speed electronic chain for the acquisition of the signals. Nevertheless, the sizes of the collimator, the sampling of the spatial distribution of the scintillation light and the thickness of the scintillator are limiting factors the spatial resolution (typically 0.3-0.4 cm). Besides, the sum of all PMT signals is used to perform the energy discrimination necessary in rejecting events that have undergone Compton scattering in tissue and would otherwise contribute noise to the image (spurious events). Therefore, gamma-cameras are photon-counting detectors (maximum count-rate of about 300.000 events/second) requiring a good energy resolution and fast scintillators.

Being the gamma-camera expensive and bulky, there is always a lot of interest to develop compact and portable tools, that allow to visualize small organs (e.g. tyroid) or to study the radiopharmaceuticals in small animals. So, new prototypes of mini-gamma cameras are developed. They are essentially based on multianode position-sensitive photomultiplier tubes (PSPMT) coupled to segmented or continuous scintillation crystals [2].

Pinhole scintigraphy, which uses pinhole collimators, allows a more elevated spatial resolution in comparison to the parallel-hole scintigraphy and also a best efficiency of detection for small subjects set near the collimator [2-6].

In general, in all scintigraphs, the spatial resolution depends on the bore of the collimator which is dominating on other parameters, so its improvement of can be attained at expense of the geometrical efficiency. A further little enhancement can be obtained using pixelated detectors, but at expense of the energy resolution.

In principle, the use of semiconductor detectors would improve the camera performance. Indeed, semiconductor detectors (e.g. CdZnTe and CdTe) have better energy resolution, which permits a better Compton scatter suppression, so compact light-weight cameras should be possible thanks to a readout electronics integrated in the detector itself. Disadvantages of semiconductor detectors include low detection efficiency, due to the lack of thickness of the semiconductor, high cost, small field of view (FOV) and deleterious effects of charge-carrier trapping on energy spectra [6].

One way to overcome the conflict between spatial resolution and sensitivity, in the pinhole scintigraph, may be the use of the rotating-slit collimator [7-12]. In principle, each source-point emits continuously gammas, so its final image can be obtained through a longitudinal slit-collimator and a one-dimensional strip-detector put in parallel with it, then rotating the system in more positions around the common axis and decoding the “back-projected” images as in a CAT reconstruction (Fig.2).

The major advantage of a rotating-slit collimator is its higher acceptance solid-angle to gamma flux than the pinhole collimator. In any case, the rotating slit is not quite investigate, above all in its possibility to reject the spurious events regardless of the energy resolution and in the use of a bundle of scintillating fibers, put in parallel with the slit, which permit the avoiding of the parallax error by a cross-area readout, using an integrating image detector.

The today's possibility to produce scintillating glass fibers with elevated "light yield" and the availability of CCD systems coupled with compact image intensifiers, allow us to test innovative technology on a rotating-slit scintigraph, able to overcome the actual limits of the pinhole and parallel-hole scintigraphy.

A further application of the rotating-slit collimator is its off-axis version [13] which permits to attain tomographic information if better decoding algorithms are developed to erase unwanted blurred planes.

## 2. Imaging with a rotating-slit collimator and spurious rejection

A slit collimator works in such manner that a point source casts a shadow, on a plane parallel to that of the slit, in the form of a straight line oriented in the same direction of the slit so to obtain one “back-projection”. Fig.1 shows the geometry of the system proposed by us, where a one-dimensional plane detector, put crosswise the incoming gammas, in parallel to the slit and forming one piece with it, rotates around its axis to achieve the so-called “back-projection image” from which the final image can be reconstructed. Under this respect, a 2D detector should be quite redundant since the valid coordinates are only in the direction of the slit.

The width of the slit is determinant for the spatial resolution of the scintigraph, while it is convenient that its length be as large as possible to gain geometrical efficiency.

Essentially, in scintigraphy is the *functional activity* of the activated organ which acquires relevance. Indeed, the radionuclides carried by the radiopharmaceuticals are above all absorbed by the organ of interest so they emit gammas continuously and regularly from the same source. Hence, the final image can be aptly reconstructed from the back-projections achieved in all the rotation angles, while the gamma emitted randomly, so interesting only one back-projection each, are avoided forming an uniform background.

About, the possible segmentation of the detector this would deteriorate its energy resolution, especially using long scintillating strips, due their probable disuniformity, cross-talk and light absorption. But, in our case, the spurious events are avoided without energy analysis, that is in the process itself of image reconstruction.

Another problem concerns the counting rate limitation, which requires fast scintillators and fast readout system, if the photon counting procedure is adopted. Now, with a slit collimator, the incoming rate of the events is also two order of magnitude greater than other collimators, so the adoption of an integrating detector (CCD), which does not limit the acquisition rate, of the events, appears the just solution, given that in our case the energy analysis is not necessary.

Fig.3 shows the mathematical procedure of reconstruction of the scintigraphic image $f^*(\theta,r)$. Indeed, the back-projected image $b(\theta,r)$ can be considered derived from the image of departure $f(\theta,r)$ convoluted with the function $\pi/r$, which represents the back-projected impulsive answer of the system. As it is known, the operation of convolution becomes a simple product in the domain of the Fourier transform, so the inverse transform of the Fourier transform of $b(\theta,r)$ divided the transform of $\pi/r$ produces the final image.

Fig.2 shows the various types of events involved in a scintigraphic process. Obviously, Compton or other background events cannot be reconstructed, due their random appearance in space and in time.

In conclusion, if the photon-counting procedure of acquisition includes the energy analysis of the events, this analysis becomes not necessary in a rotating-slit scintigraph. Hence, a rotating-slit scintigraphy permits the use of integrating detectors, as the CCDs, so scintillators with a decay time of the order of milliseconds can be employed, just as the Tb-doped scintillating glass which can be stretched easily in fibers.

The spurious rejection capability of a rotating-slit collimator has been tested with Montecarlo simulations as well as experimentally. Fig.4 represents the image reconstruction of two gamma sources (2 mm diameter) using 100 back-projections (1000 gammas per projection) without spurious events and with a slit 1 mm width. Fig.5 represents still the reconstruction of the same gamma sources, using 100 back-projections (1000 gammas per projection), but introducing 10 random events per each gamma. Therefore, the spurious rejection capability appears robust, if these events can be eliminated also when their number is superior of an order of greatness to the valid events. It is interesting to note an energy analysis of the signals can eliminate only Compton events, while the rotating-slit procedure can avoid also other background gammas.

## 3. Parallax error elimination

The knowledge of the depth of interaction of the gamma rays is important in pinhole imaging, because the gammas can intersect the detector face at a large angle and introduce a parallax error, with consequent degradation of the spatial resolution. Therefore, the transversal 2D readout of the bundle of scintillating fibers (Fig. 6) can solve of the conflict between the minimization of the parallax error and the improvement of the detection efficiency, permitting the high absorption thickness of the scintillator.

The position of impact of the gammas can be determined with precision thanks to the high segmentation of the detector, possible with the use of scintillating fibers and their transversal 2D readout. Essentially, the contents of the pixels of the cross-image, lined up with the slit within a small angular interval, are summed to obtain an 1D image for every angle of rotation. Then, such 1D images are back-projected and added for the reconstruction of the final image.

Fig.7 shows the schematic drawing of the longitudinal view of the detector with two different sizes of the slit: note the correlation between the length of the slit and the length of the bundle of the scintillating fibers.

## 4. The scintillating-glass fiber-detector

Scintillators are commonly used to convert the energy of gammas in visible photons, but their segmentation in fibers of size of the order of one millimeter and lengths of at least 100 mm is problematic, unless we use scintillating glass fibers.

Fig. 8 shows the two chances of positioning of the same bundle of fibers in respect to the axis of rotation. If the axis of rotation is set on the edge of the bundle it is possible to double the diameter of the FOV.

About the texture of the fiber bundle, after many tests, the highest collection efficiency of the scintillation light is reached using air as cladding. Therefore, we propose the structure showed in Fig.9, where piles of uncladded fibers are divided by sheets of dark plastic which acts as extra mural absorber (EMA) to absorb the non-guided light.

We use fibers of scintillating silicate glass, terbium doped, barium charged, named LKH-6 [14] (effective atomic number $Z_{eff} \cong 30$, maximum of emission ≅ 550 nm, decay time = 3÷5 ms, light yield ≅ 40 ph/keV, attenuation length at 140 keV ≅ 5 mm). In this case, the use of air as cladding permits to improve the collection efficiency of the scintillating light, in one direction, from 3.46 % to 18.55 % [15].

The decay time of LKH-6 glass matches well with the intensified CCD read-out, being acceptable an integration time of few seconds for each rotation angle, using the typical doses of radionuclide adopted in nuclear medicine.

The CCD intensification can be obtained coupling at its input, by fiber optics, a compact MCP image intensifier. Another solution can be based on an electron-multiplying CCD (EMCCD) [16], without external intensifier, coupling the fiber bundle in front the CCD, while the charge signal is multiplied until $10^3$ within the CCD.

With reference to Fig.6 and using a known formula [4], we can see in Tab.I the spatial resolution of the scintigraph reached with a slit of effective width $d_{eff}$ = 1 mm varying size of the fibers (detector resolution ) and the distance source-slit, for two diameters of the source, being set the distance slit-detector (L=70 mm).

It is interesting to note in Tab.I as the spatial resolution of the scintigraph is practically independent on the detector resolution, but strongly dependent on the size of the slit, so that it is unhelpful, for $d_{eff}$ =1 mm, to force the resolution of the detector under 0.5 mm.

**SPATIAL RESOLUTION**

| **Detector resolution (mm)** | **0.5** | **1** | **2** |
|---|---|---|---|
| Source diameter 50 mm | 1.36 – 2.1 | 1.4 – 2.3 | 1.5 – 2.9 |
| Source diameter 25 mm | 1.18 - 1.5 | 1.2 – 1.6 | 1.23 – 1.8 |

**Tab.I**

## 5. Experimental results

Fig. 10 shows the scheme of a simple prototype of our scintigraph, absolutely portable, where we have preferred to rotate the gamma source rather than the system detector-slit. A source of gammas consisting of ~ 7cc of physiological solution, loaded with 99mTc, is contained in a cylinder of glass (2 mm thick, 22 mm external diameter). The initial activity of the gamma source is ~ 7 mCi, so the activity at the output of each of the two asymmetric holes (Fig.10 and following), practiced on a lead slab, and used as secondary sources, has been valued about 1.25 μCi, comparable with that produced by 1 $mm^3$ of biological tissue typically activated (around 105 Bq). The bundle of scintillating glass fibers (uncladded, 0.5 mm diameter, 110 mm length) was realized in collaboration with the "Stazione Sperimentale del Vetro" (SSV) of Murano-Venice.

Fig.11 shows a single side-view (with spurious) of the two gamma beams springing from the two asymmetric holes (exposure time 1s).

Fig.12 shows, after the correction of the parallax error, the image of the two gamma sources using 18 scans, with one second of exposition per every rotation angle, while Fig.13 and Fig.14 show the same image, but thresholded to avoid the steady background due to the spurious contribution.

In conclusion, as it regards the efficiency and therefore the run-time of one scintigraphy, the use of one slit (instead of one hole) increases the useful solid angle of two orders of magnitude (length of the slit about 100 times its width), while only 8 projections would be enough to have an image of good quality, as experimentally verified.

At least, as comparison, we can see in Fig.15 the image of the same two gamma sources obtained, at parity of conditions, from the commercial [17] parallel-hole scintimammograph MAMMOCAM 1000 (35 mm lead collimator depth, 1.8 mm diameter holes, pixellated CsI(Tl), 2x2 $mm^2$ pixel size, PSPMT read-out, 36 s exposure time). Note in Fig. 14 as the spatial resolution of te detector is limited by the sizes of the parallel-hole.

## 6. Conclusions

The first results obtained with the described prototype of rotating-slit scintigraph, using a scintillating glass fiber detector, learn to us that it is possible to overcome the present limitations of the gamma cameras under various aspects such as:

1. the improvement of the detection efficiency of two orders of magnitude, thanks to the high acceptance solid-angle presented to the incoming gammas and to the possible high absorption volume of the scintillator;
2. the avoiding of the parallax error, by the 2D transverse readout of the bundle of scintillating optical fibers which are put in parallel with the slit. This permits the use of scintillators with low stopping power and high absorption thickness without to limit the spatial resolution;
3. the rejection of the spurious events, without energy analysis, due the avoiding of random Compton events, and other background, for back-projection procedure.
4. the unlimited count rate of the detected events with the adoption of integrating imaging devices, which permit the use of low decay-time scintillators, such as the scintillating glasses suitable to be stretched in fibers.

**Acknowledgements**

The authors are indebted with Prof. U. Mazzi of the Department of Pharmaceutical Sciences, University of Padua, for his collaboration to perform tests with Technetium-99m.

**Figure captions**

Fig.1. Rotating-slit scintigraph set-up.
Fig.2. Schematic drawing of back-projection imaging and possible events.
Fig.3. Reconstruction procedure of the final image from the back-projections and spurious rejection.
Fig.4. Image reconstruction, by Montecarlo simulation, of two gamma emitting holes (2 mm diameter) from 100 back-projections (1000 gammas per projection), without spurious events.
Fig.5. Image reconstruction, by Montecarlo simulation, of two gamma emitting holes (2 mm diameter) from 100 back-projections, random introducing spurious events.
Fig.6. Schematic drawing of the detector cross-section showing the possibility of the parallax-error rejection.
Fig.7. Schematic drawing of the longitudinal view of the detector with two different sizes of the slit.
Fig.8. Fiber bundle read-out with intensified CCDs
Fig.9. Structure of the used scintillating-glass fiber bundle.
Fig.10. Schematic drawing of the experimental set-up.
Fig.11. One side-view (with spurious) of the two gamma beams from the two asymmetric source holes (exposure time 1s).

Fig.12. Image of the two 1 mm-diameter gamma sources (18 projections, 1s exposure/projection).
Fig.13. Thresholded gamma image of the two 1 mm-diameter gamma sources (18 projections, 1s exposure/projection).
Fig.14. Gamma image of the two 1 mm-diameter gamma sources using the parallel-hole scintimammograph MAMMOCAM 1000.
Fig.15. 3D image of Fig.13.

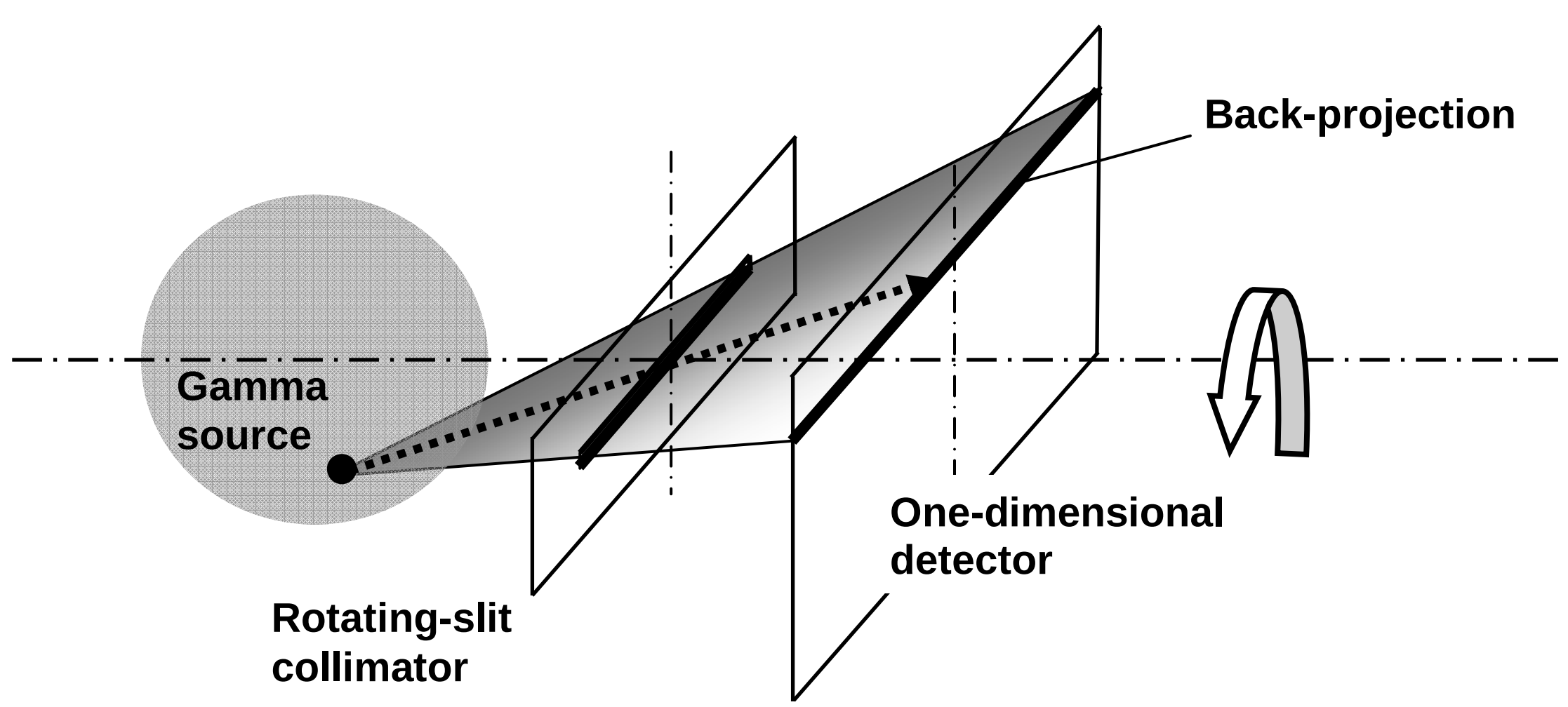

**Fig. 1**

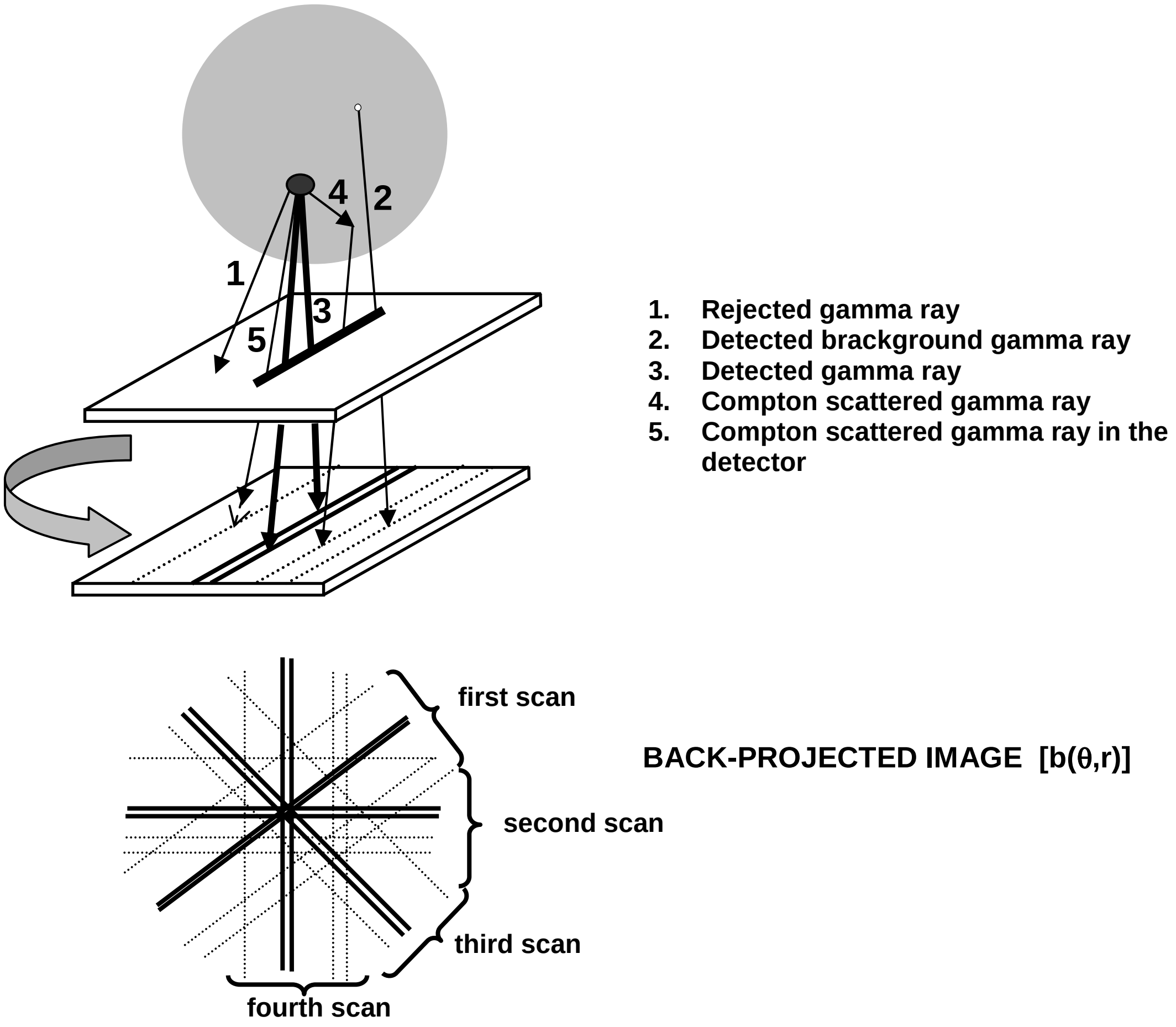
4
2
1
3
5
1. Rejected gamma ray
2. Detected brackground gamma ray
3. Detected gamma ray
4. Compton scattered gamma ray
5. Compton scattered gamma ray in the detector
first scan
BACK-PROJECTED IMAGE [b(θ,r)]
second scan
third scan
fourth scan

**FIG. 2**

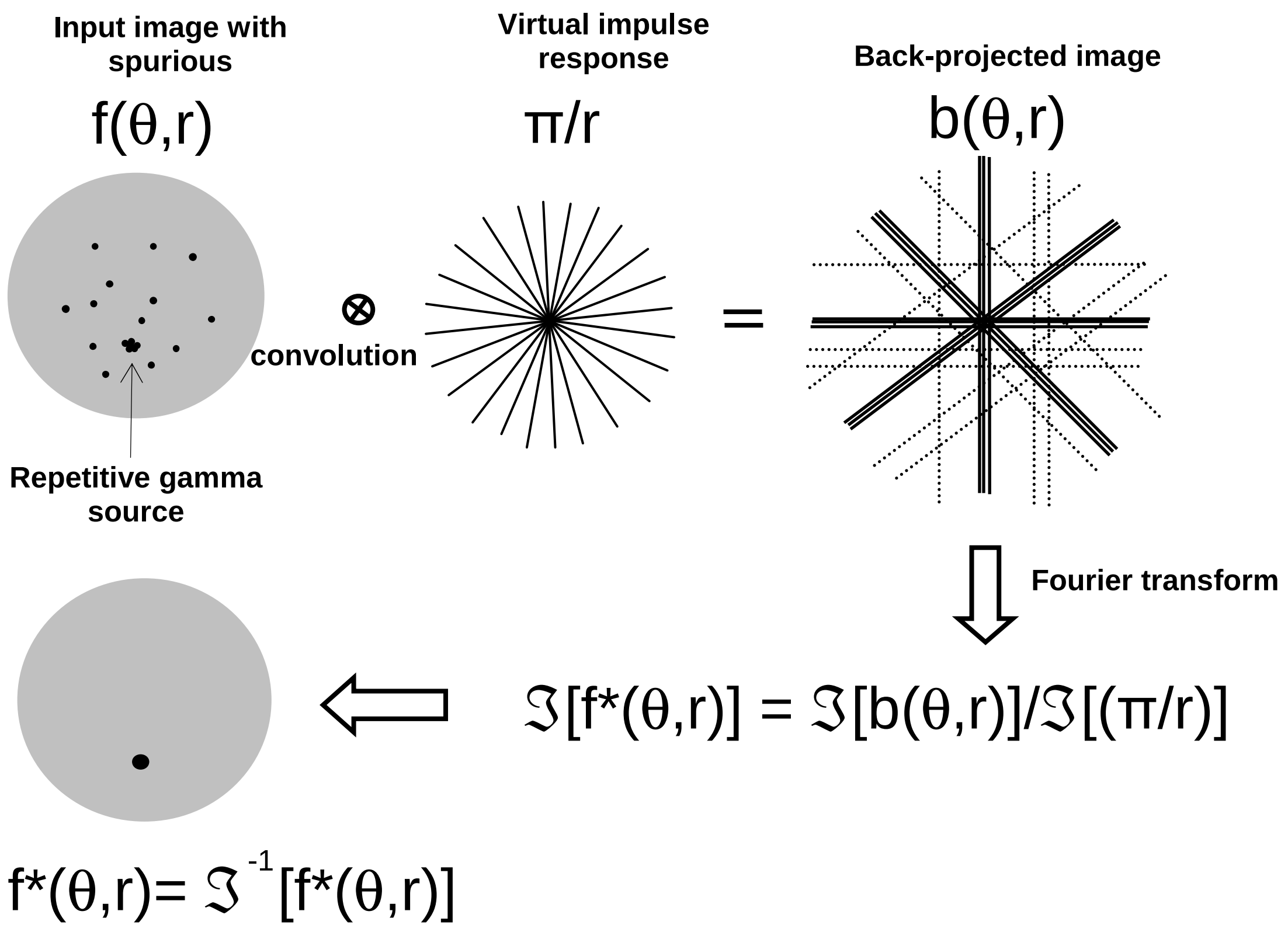

**FIG. 3**

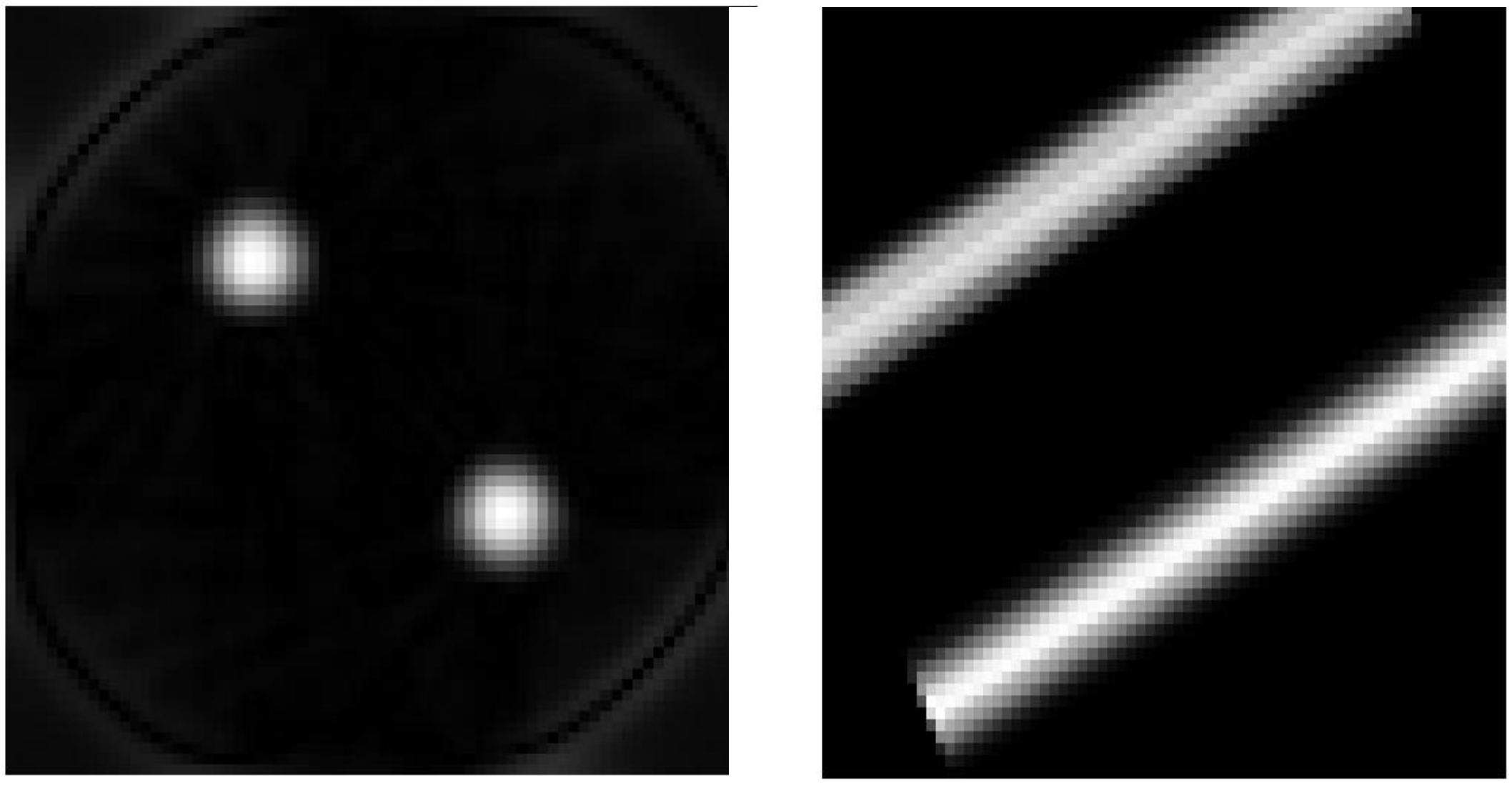

2 mm diameter gamma sources

One back-projection

**FIG. 4**

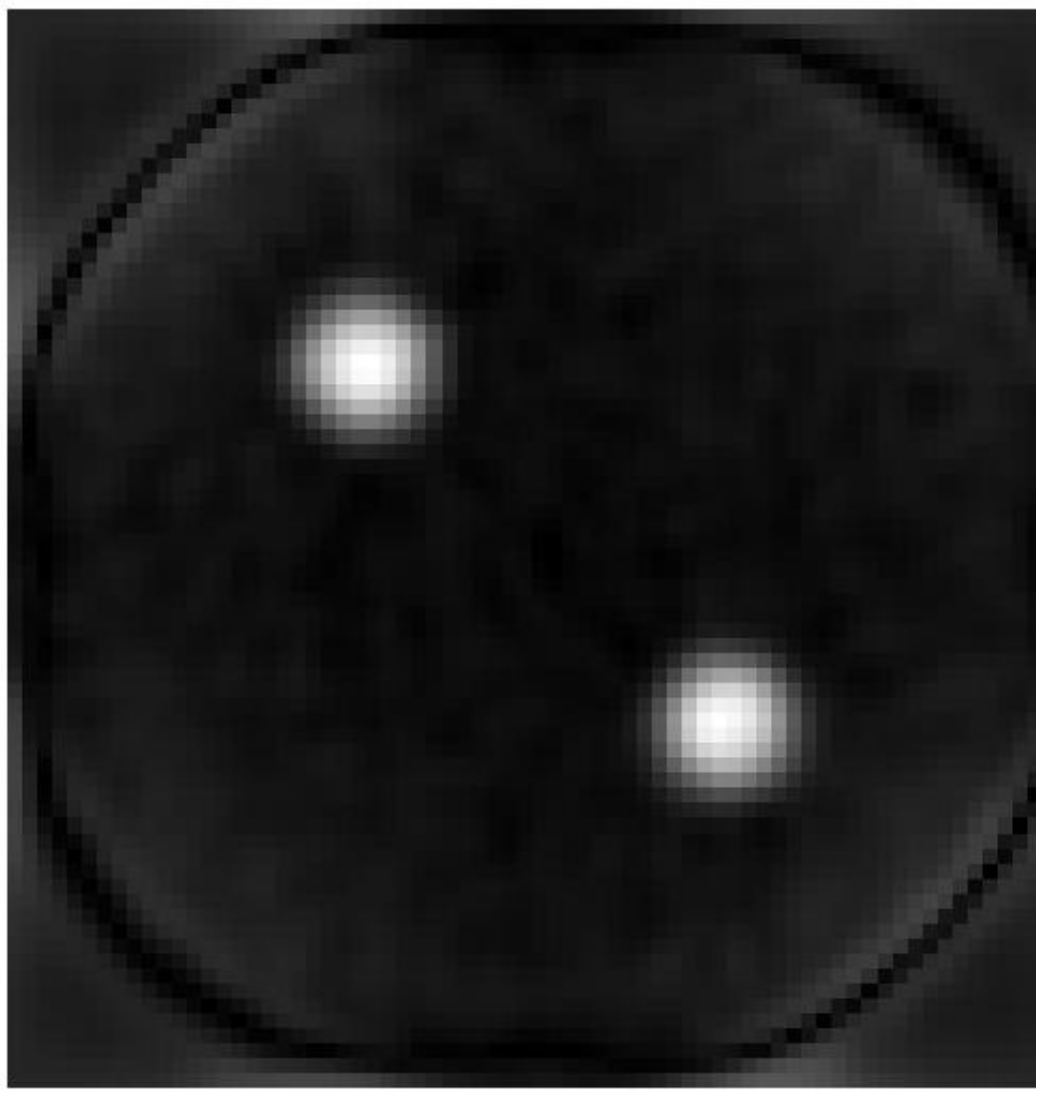

2 mm diameter gamma sources

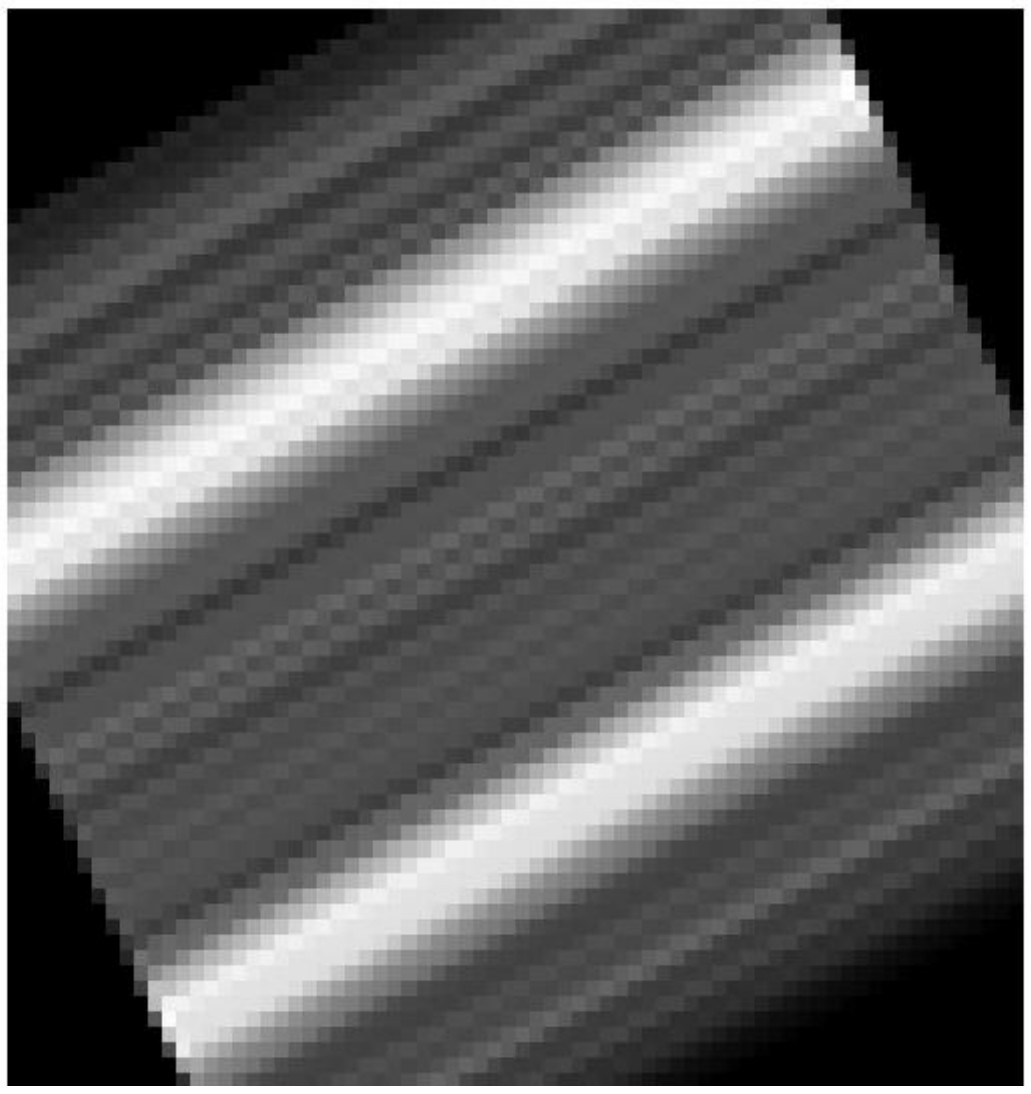

One back-projection

**FIG. 5**

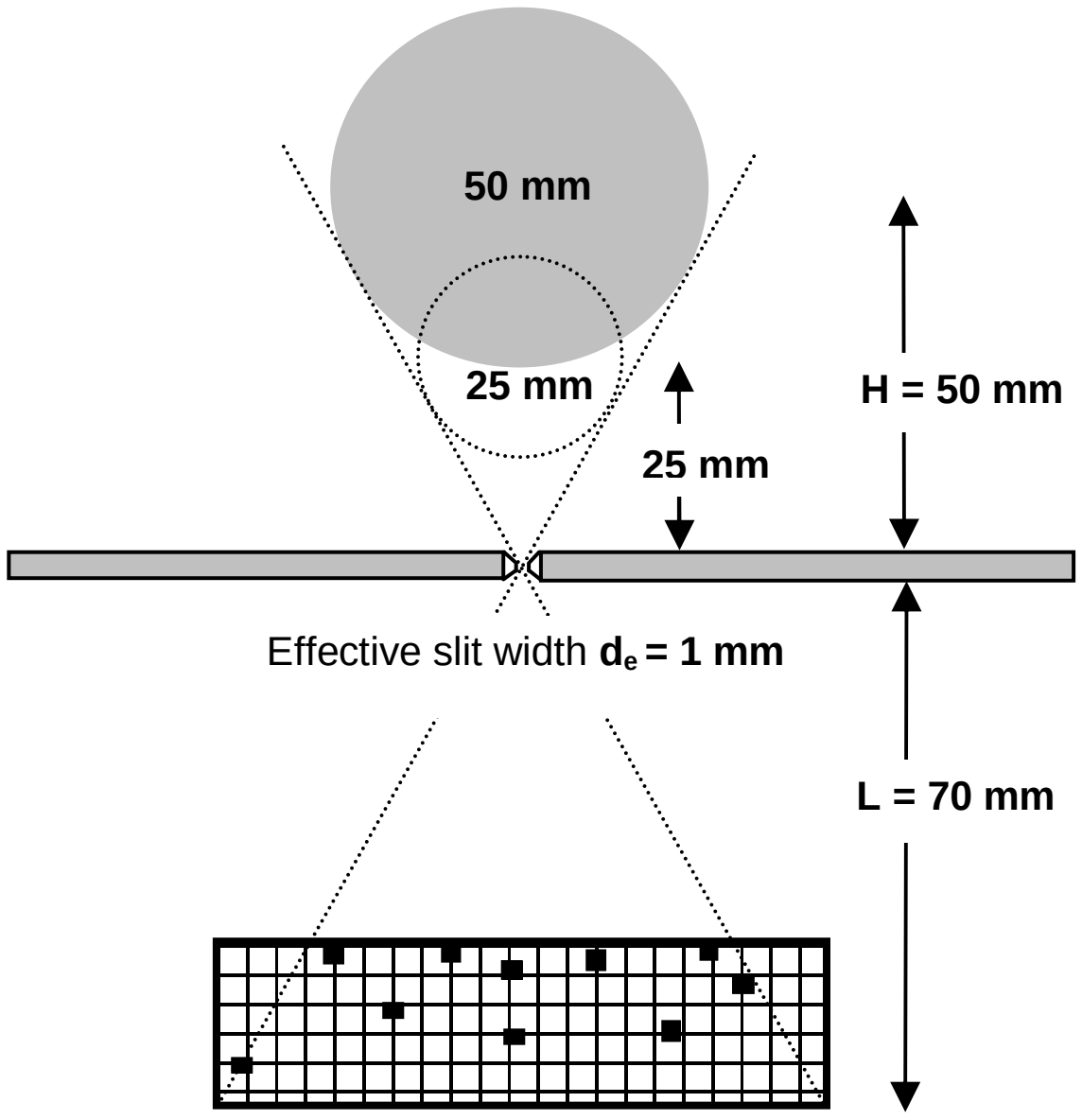

**FIG. 6**

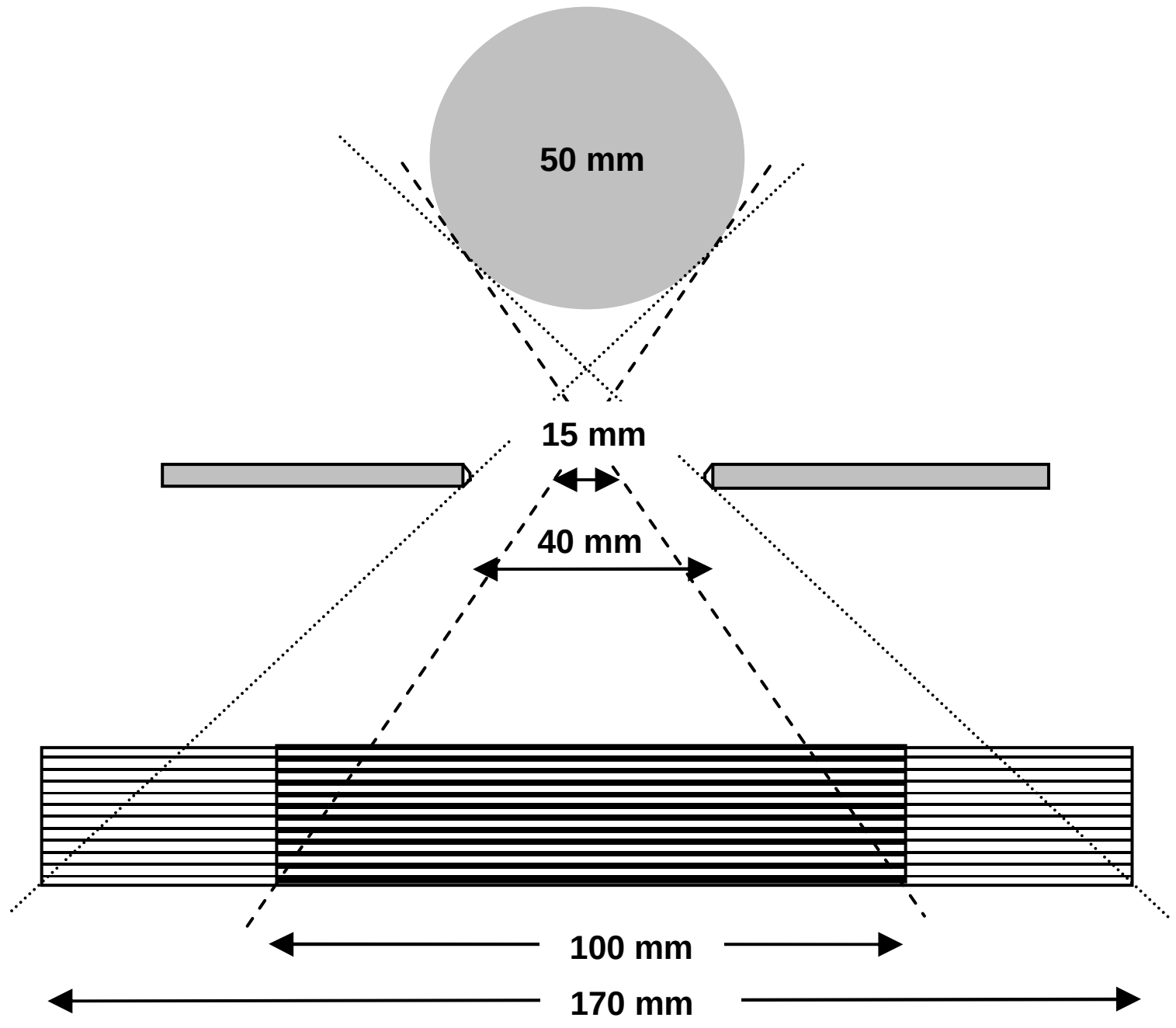
50 mm
15 mm
40 mm
100 mm
170 mm

**FIG. 7**

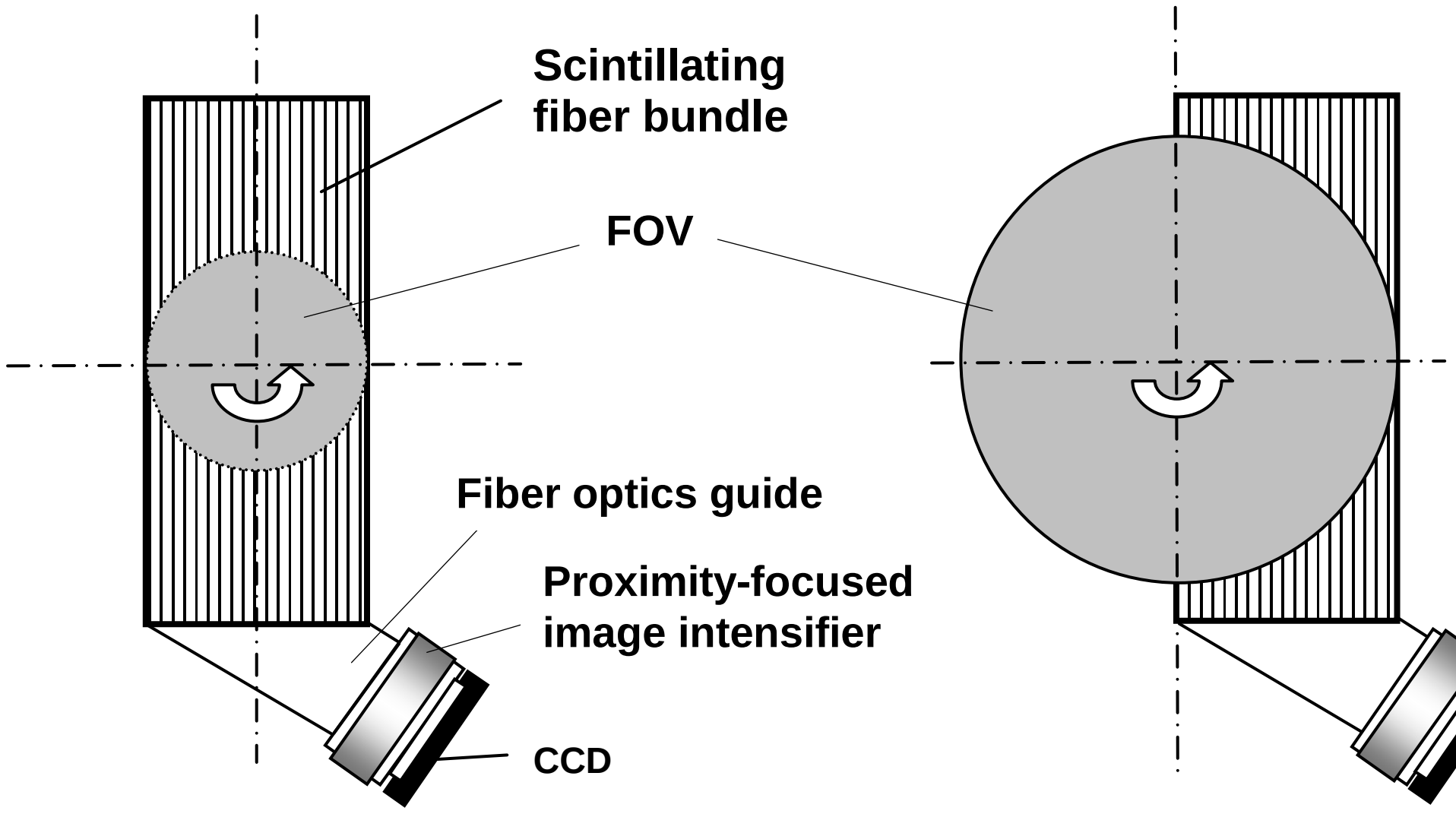

**FIG. 8**

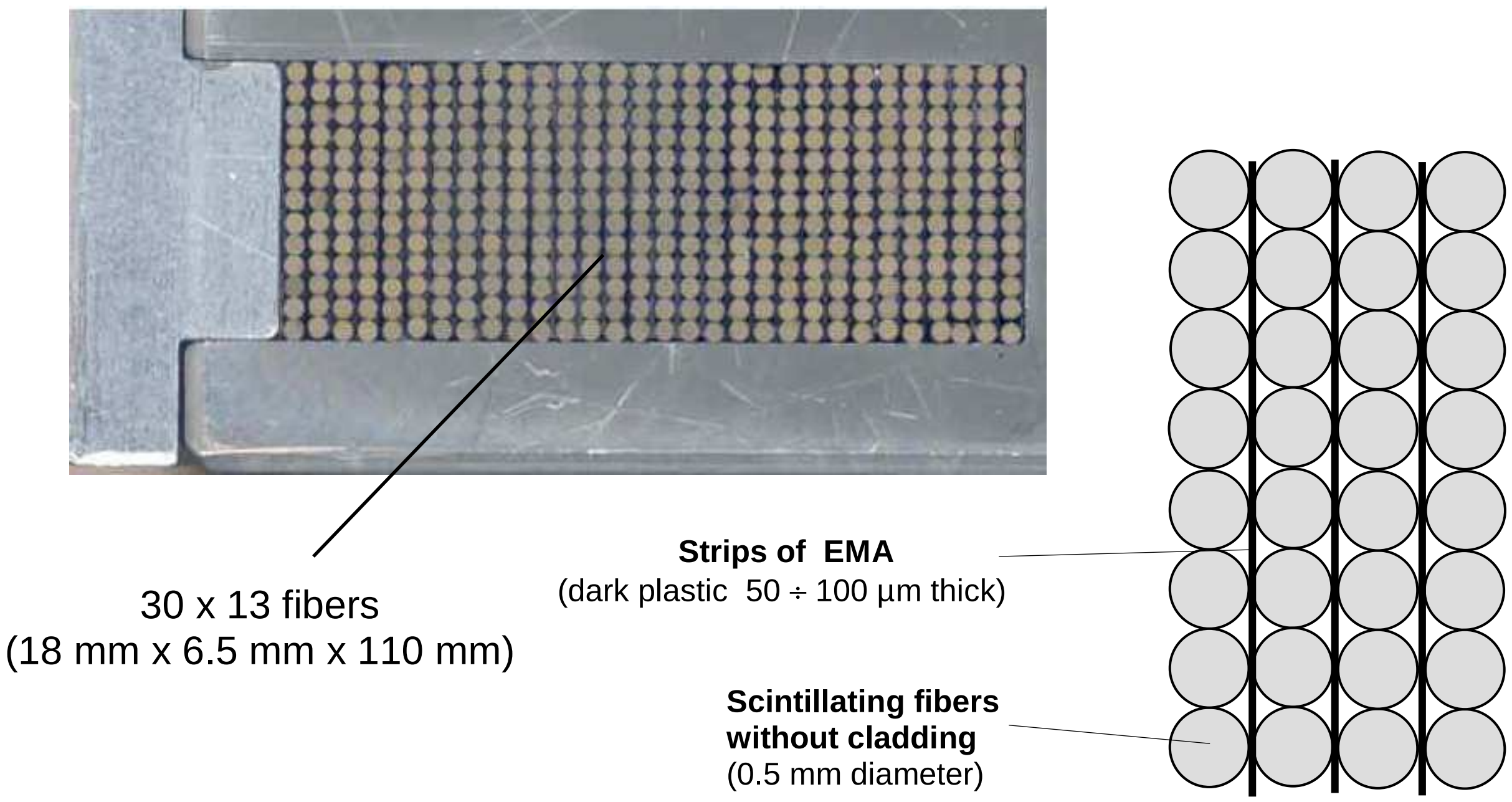
30 x 13 fibers
(18 mm x 6.5 mm x 110 mm)
Strips of EMA
(dark plastic 50 ÷ 100 µm thick)
Scintillating fibers
without cladding
(0.5 mm diameter)

**FIG. 9**

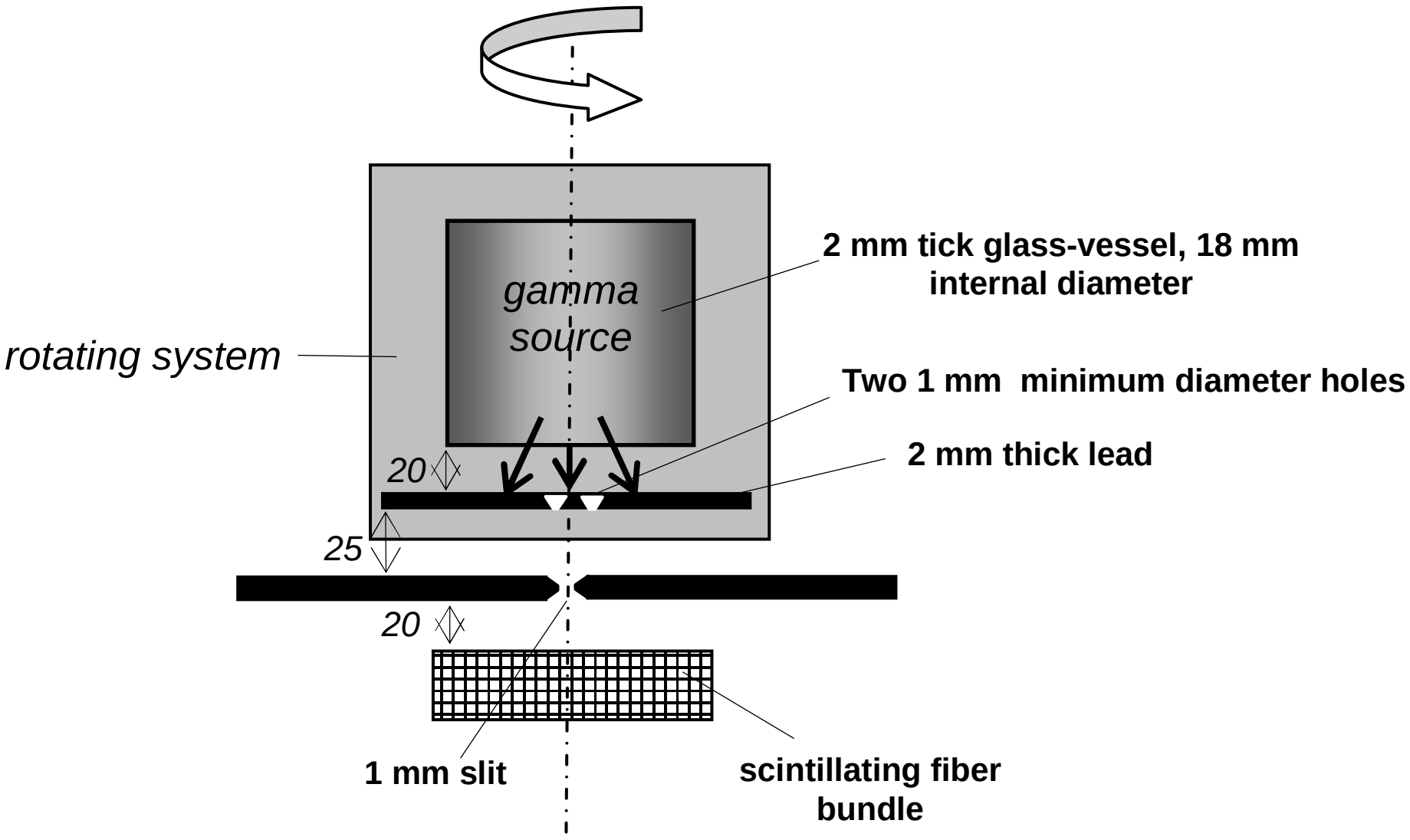
gamma
source
rotating system
2 mm tick glass-vessel, 18 mm
internal diameter
Two 1 mm minimum diameter holes
2 mm thick lead
20
25
20
1 mm slit
scintillating fiber
bundle

**FIG. 10**

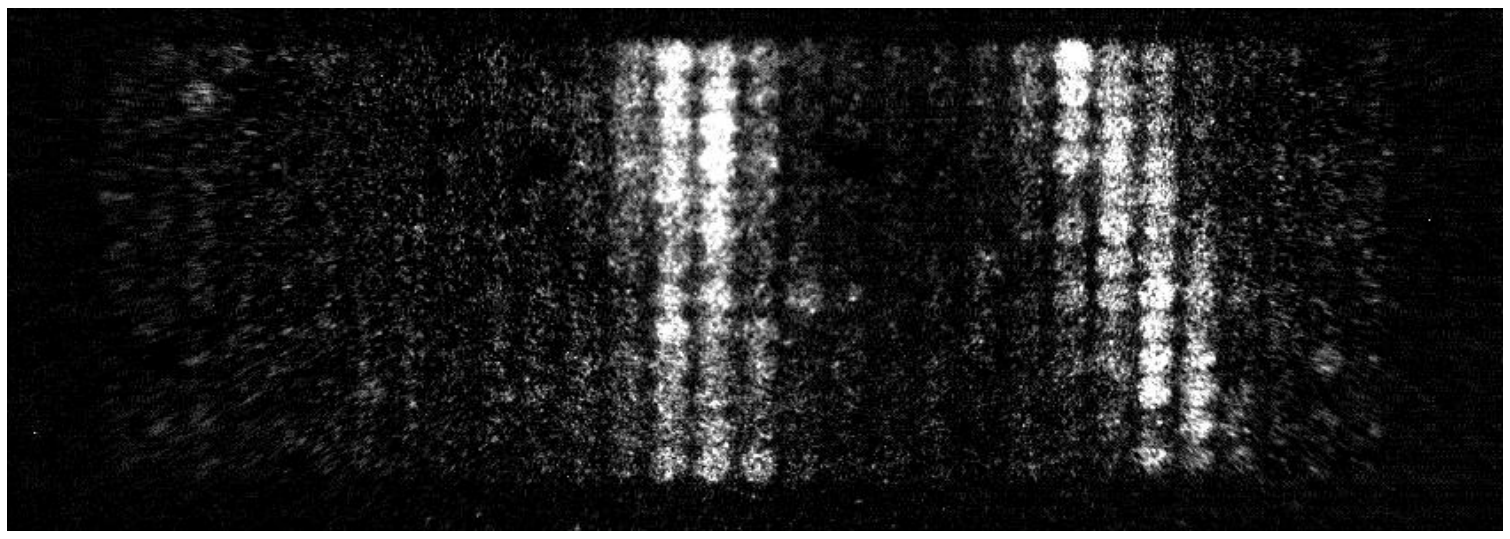

FIG.11

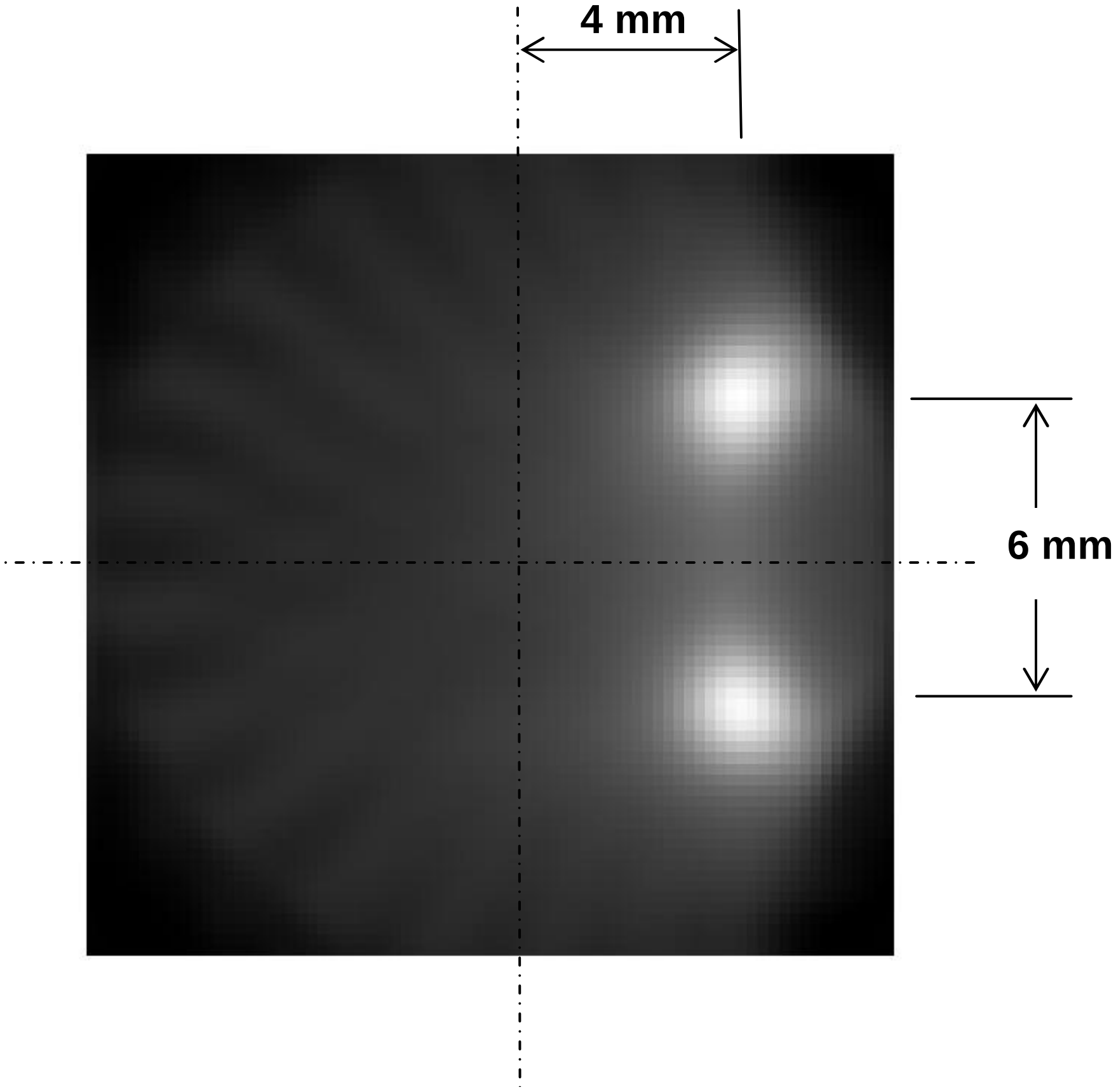

FIG. 12

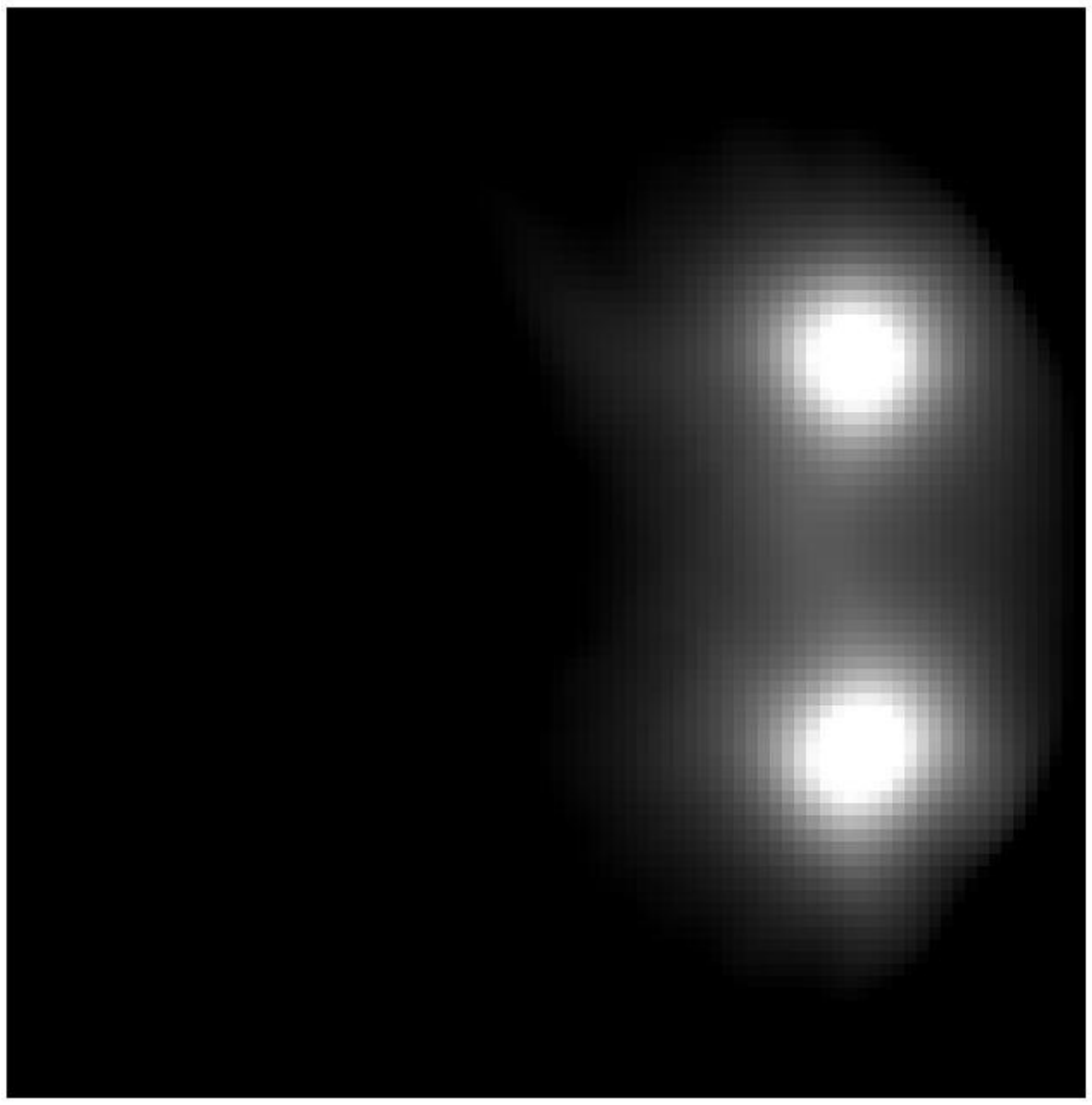

**FIG. 13**

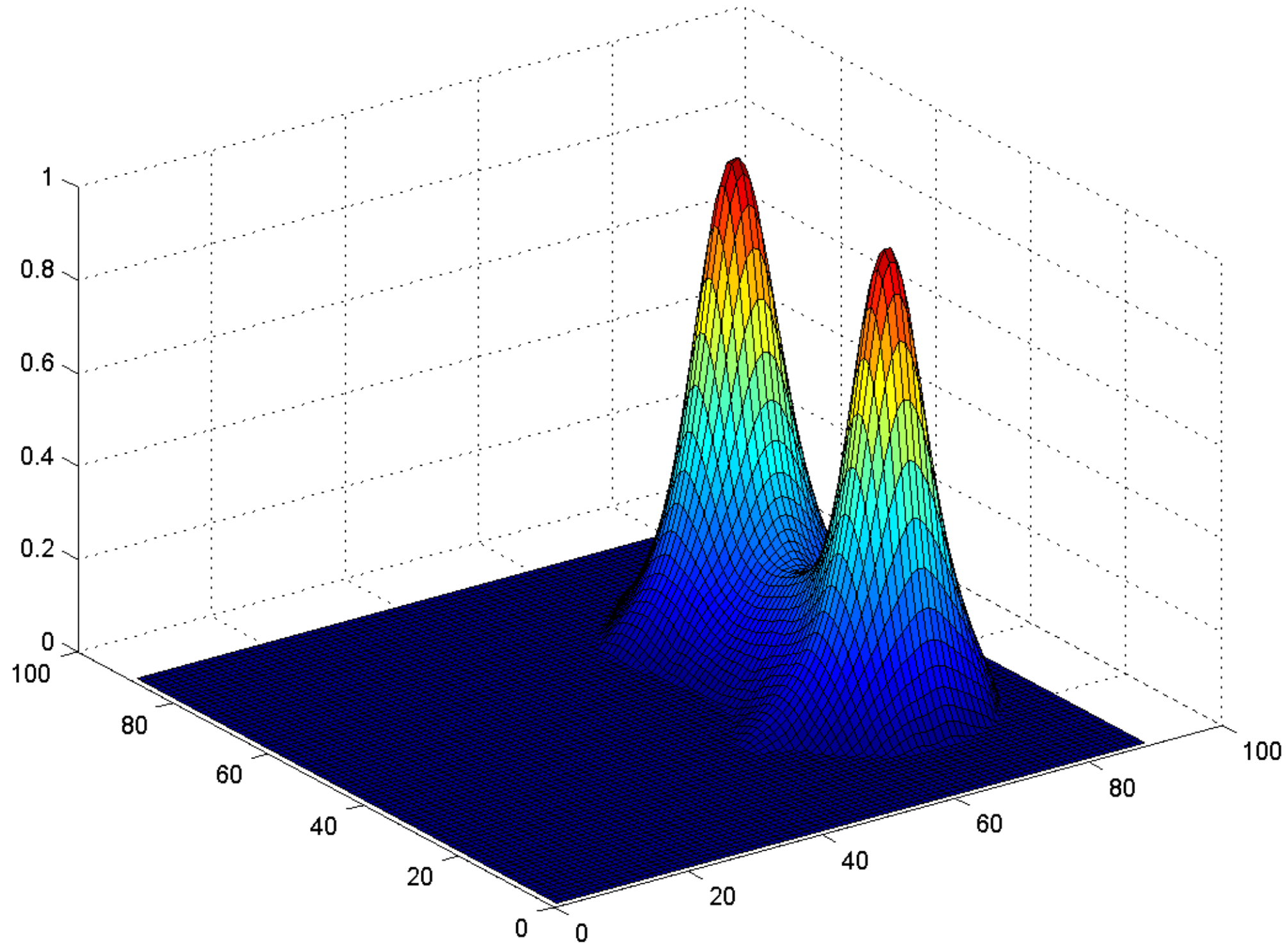

**FIG. 14**

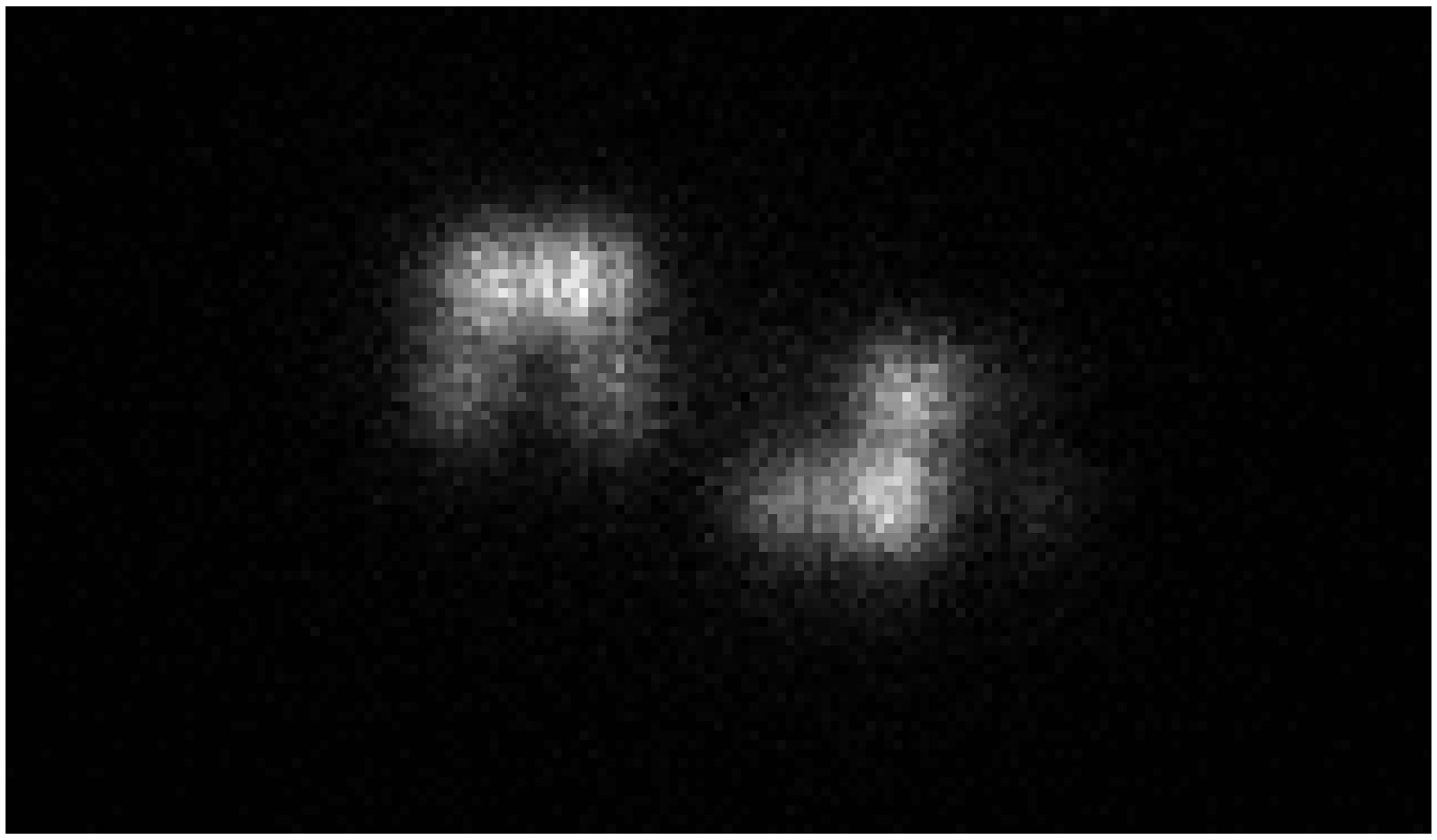

**FIG. 15**